\documentclass[%
reprint,
amsmath,amssymb,
prb,
]{revtex4-2}

\usepackage{graphicx}
\usepackage{dcolumn}
\usepackage{bm}
\usepackage{subfigure}
\usepackage{booktabs}
\usepackage{mhchem}
\usepackage[colorlinks=true, letterpaper=ture, pdfstartview=FitV, linkcolor=blue, citecolor=blue, urlcolor=blue]{hyperref}

\begin{document}

\preprint{APS/123-QED}

\title{Role of Characteristic Length Scale in Interface Graphitization-Induced Wear Resistance of Diamond and Amorphous Carbon}

\author{WenLiang Shi$^{1,2}$}
\author{YanYu Zhang$^{3}$}
\author{WenZheng Liao$^{1,2}$}
\author{Kai Xu$^{1}$}
\author{HongYu Wu$^{4}$}
\thanks{wuhy@szlab.ac.cn}
\author{ZhiCheng Zhong$^{4,5}$}
\author{KeKe Chang$^{1}$}
\thanks{changkeke@nimte.ac.cn}

\affiliation{\\$^1$State Key Laboratory of Advanced Marine Materials, Zhejiang Key Laboratory of Extreme-environmental Material Surfaces and Interfaces, Ningbo Institute of Materials Technology and Engineering, Chinese Academy of Sciences, Ningbo, Zhejiang 315201, China}

\affiliation{$^2$College of Materials Science and Opto-Electronic Technology, University of Chinese Academy of Sciences, Beijing 100049, China}

\affiliation{$^3$CAS Key Laboratory of Magnetic Materials \& Zhejiang Province Key Laboratory of Magnetic Materials and Application Technology, Ningbo Institute of Materials Technology and Engineering, Chinese Academy of Sciences, Ningbo 315201, China}

\affiliation{$^4$Suzhou Lab, Suzhou 215123, China}

\affiliation{$^5$Suzhou Institute for Advanced Research, University of Science and Technology of China, Suzhou 215123, China}


\date{\today}

\begin{abstract}
The evolution of interfacial atomic structures critically influences the friction and wear behavior of carbon-based materials. However, how the characteristic length scale of friction-induced sp² reconstruction governs macroscopic wear remains poorly understood, particularly for diamond and amorphous carbon where the interfacial graphitization modes differ fundamentally. In this work, we develop a machine learning potential for these carbon systems and investigate the structural evolution at interfaces in both diamond/diamond and amorphous/amorphous carbon systems using molecular dynamics simulations. Our results reveal distinct atomic-scale characteristics of graphitization at the two interfaces. Diamond interfaces develop a laterally continuous sp\textsuperscript{2} reconstruction layer with a characteristic length of 30--45~\AA, while amorphous carbon interfaces form only fully isolated sp\textsuperscript{2} patches of 8--12~\AA. This disparity in characteristic length scale determines the density of weakly bonded interfacial atoms left outside the reconstruction layer, thereby directly dictating the macroscopic wear rate. Based on these insights, we propose a strategy to regulate friction-induced graphitization in diamond coatings by protecting specific crystallographic orientations, such as the (111) close-packed planes. This work bridges the gap between atomic-scale interfacial structure and macroscopic tribological performance, offering mechanistic guidelines for the rational design of wear-resistant carbon-based coatings.
\end{abstract}

\maketitle


\section{Introduction}
Friction and wear are fundamental physical processes that critically govern the reliability and service life of precision machinery and protective coatings \cite{Wong2016,Blau2001,Perry2005,Tambe2005,Tambe2005a}. From the perspective of microscopic mechanisms \cite{Erdemir2018}, wear originates from the migration, adsorption, desorption, and structural reconstruction of interfacial atoms driven by external forces \cite{Markov2022}. These atomic-scale dynamic behaviors govern interfacial structural evolution and ultimately determine macroscopic friction and wear performance \cite{Mueser2001,Zhou2023}. Yet identifying the characteristic structural length scale that bridges atomic-level interfacial reconstruction and macroscopic wear rate has remained a long-standing conceptual challenge, particularly for carbon-based materials where friction-induced phase transformation complicates the structural picture. Unraveling this microstructural origin is therefore essential for the rational design of high-performance wear-resistant materials \cite{Carpick1997,Zhai2021}. 

The microscopic structural evolution pathways are strongly dependent on the nature of the contacting interface, and differences in this pathway can lead to pronounced variations in macroscopic wear performance \cite{Filleter2009,Gao2024}. Carbon-based materials provide a compelling example: both diamond and amorphous carbon are widely employed as advanced wear-resistant coatings owing to their exceptional mechanical and tribological properties \cite{Erdemir2006}, yet diamond generally exhibits superior wear resistance compared to amorphous carbon \cite{Liang2024}. Notably, extensive experimental studies have confirmed that friction induces graphitization at both types of interfaces \cite{Zhou2023,Li2020}, generating graphite-like structures that play a dominant role in governing interfacial tribological behavior \cite{Li2020,Lin2022}. This shared phenomenon of friction-induced graphitization thus provides a critical structural basis through which the contrasting wear behaviors of diamond and amorphous carbon can be examined.

Despite this recognition, several fundamental scientific questions remain unresolved. The precise microscopic characteristics of friction-induced graphite-like structures are still unclear \cite{Liu2020,Pastewka2011}, and the dynamic evolution of these structures during the friction process has yet to be tracked at the atomic level \cite{Erdemir2006, Erdemir2001,Konicek2012,Miyoshi1992}. More importantly, it remains unexplained why the graphitization behaviors of these two materials differ, and how these differences ultimately give rise to the distinct macroscopic wear rates observed experimentally \cite{Li2018a,Mo2009,Wang2018}. In particular, whether the characteristic length scale of the friction-induced sp\textsuperscript{2} reconstruction layer differs fundamentally between diamond and amorphous carbon interfaces, and whether such a difference constitutes the governing structural parameter for macroscopic wear, has never been systematically addressed. A key bottleneck is that experimental characterization techniques currently struggle to provide real-time, atomic-resolution monitoring of interfacial structural evolution during friction.
 
\begin{figure*}[htbp]
\centering
\includegraphics[width=0.8\textwidth]{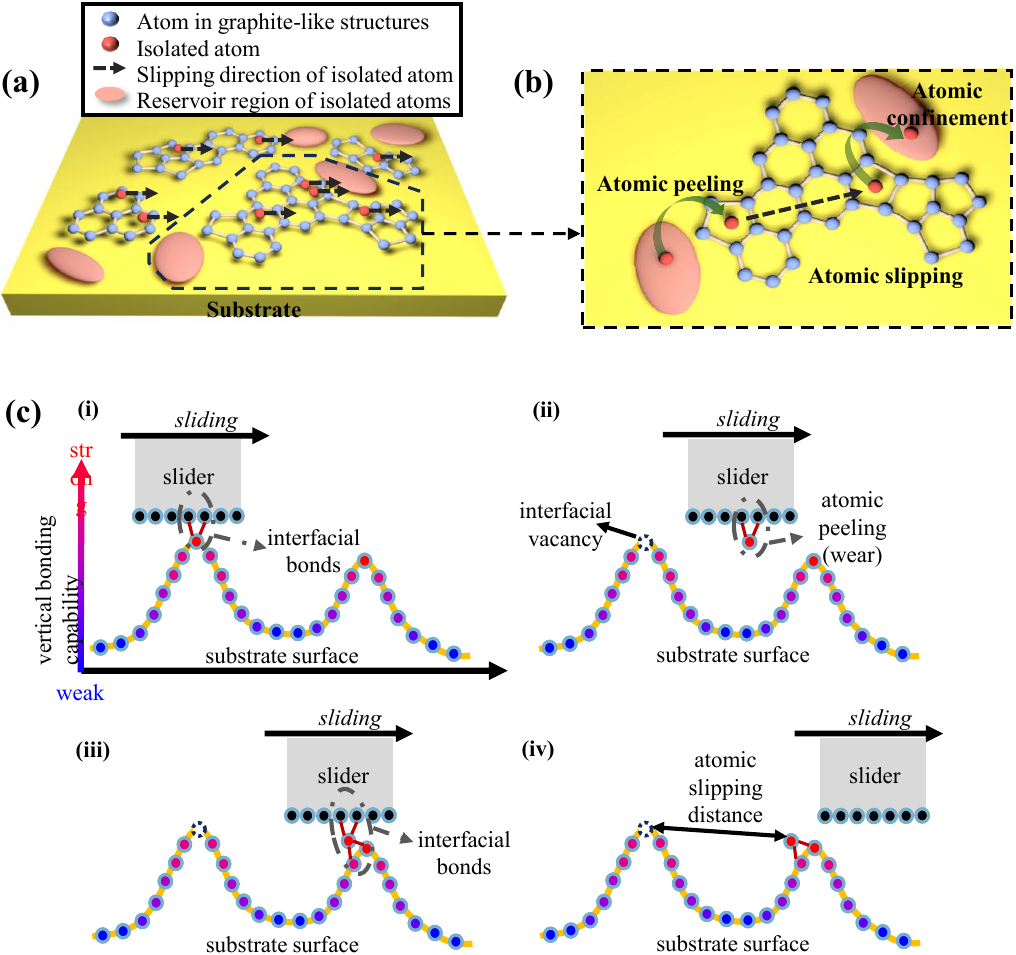}
\caption{Atomic-scale structural characteristics and wear mechanisms at graphitized carbon interfaces. (a) Schematic illustration of the local atomic structures and representative atomic motions at the graphitized surface of a carbon-based material during friction. (b) Graphite-like structures formed at the surface create local regions of weakened out-of-plane bonding, which facilitate the peeling and lateral slipping of isolated interfacial atoms. (c) Spatial heterogeneity in the out-of-plane bonding strength within the local surface structure of the substrate enables atomic slipping through dynamic bond-breaking and bond-forming processes.}
\label{fig1}
\end{figure*} 

Atomic-scale simulation offers a powerful route to overcome this limitation. Molecular dynamics (MD) simulation, in particular, enables direct tracking of atomic trajectories and structural transformations, making it well-suited for uncovering the fundamental origins of macroscopic tribological behavior \cite{Zhou2023,SanchezLopez2003}. However, traditional MD relies on empirical interatomic potentials. These potentials are not accurate enough to describe the complex bonding changes in carbon systems, which limits their ability to capture the formation and evolution of interfacial structures \cite{Soules1982, Erdemir2004,Kuwahara2019,Ma2009,Wang2019a}. Although density functional theory (DFT) provides electronic-structure-level accuracy, its prohibitive computational cost makes large-scale, long-timescale friction simulations impractical. In recent years, machine learning potential methods trained on high-quality DFT datasets have emerged as a compelling solution, achieving DFT-level accuracy at a fraction of the computational cost and enabling large-scale MD simulation \cite{Zhang2018,Zhang2019,Zhang2020a,Fan2022,Fan2021,Ying2025, Zhao2023,Batatia2022}. This combination of precision and efficiency makes machine learning potentials particularly well-suited for modeling complex interfacial behaviors in friction systems, effectively bridging the gap between traditional MD and first-principles DFT, with successful applications demonstrated across diverse material and interface systems \cite{Hu2022}.

In this work, we develop a machine learning potential for diamond and amorphous carbon systems and employ it in MD simulations to investigate the interfacial structural evolution during friction. Our simulations successfully reproduce the experimentally observed graphitization phenomenon at both types of interfaces and reveal a fundamental disparity in the characteristic length scale of the friction-induced sp\textsuperscript{2} reconstruction layer: diamond interfaces develop a laterally continuous reconstruction layer extending 30--45~\AA, whereas amorphous carbon interfaces form only fully isolated sp\textsuperscript{2} patches of 8--12~\AA. We establish a causal chain linking this characteristic length scale to the density of weakly bonded interfacial atoms left outside the reconstruction layer, which in turn directly determines the macroscopic wear rate. Based on these mechanistic insights, we further propose a coating design strategy that regulates the characteristic length scale of graphitization in diamond coatings by protecting the (111) close-packed planes, offering atomic-level guidance for the rational design of carbon-based wear-resistant coatings.

\section{computational methods}

Accurately describing the interatomic interactions in carbon-based friction systems is a fundamental prerequisite for revealing the atomic-scale mechanisms of wear. Carbon atoms can form a wide variety of bonding configurations, and the transitions among these bonding configurations are central to friction-induced structural evolution. Capturing such complex, dynamically varying bonding environments poses a significant challenge for conventional empirical potentials, including the Tersoff potential \cite{Tersoff1989}. This potential is widely used in carbon MD simulations \cite{Tomas2019,Tomas2016,Li2018}, whose fixed functional forms inherently limit their transferability across diverse bonding states. To address this limitation, we develop a Deep Potential (DP) machine learning potential specifically tailored for diamond, amorphous carbon, and their interfaces. The DP method constructs a deep neural network that maps the local atomic environment to the potential energy surface, with the network trained on a comprehensive dataset of DFT-level reference calculations. This approach inherits DFT-level accuracy in describing atomic forces and energies while enabling large-scale, long-timescale MD simulations at a computational cost orders of magnitude lower than that of direct DFT, which is essential for capturing the collective atomic dynamics and long-range structural evolution at frictional interfaces. Full details of the training dataset construction and model training procedure are provided in the Supplementary Materials.

The DP model achieves a root-mean-square error of 0.043 eV/atom for energies and 0.45 eV/Å for atomic forces, demonstrating excellent agreement with the DFT reference across diverse bonding environments at diamond and amorphous carbon interfaces. To rigorously validate the DP model, we construct an independent test dataset comprising DFT-calculated configurations systematically excluded from the training set, and benchmark its performance against that of the Tersoff potential. Critically, unlike the Tersoff potential, our DP model correctly reproduces the experimentally known diamond (001) surface reconstruction \cite{Kawarada1995,Petukhov1999}, which is a sensitive structural benchmark that requires an accurate description of surface behavior. This capability is particularly important for faithfully capturing the onset of friction-induced graphitization at the interface. The corresponding validation results are presented in  FIG.~\ref{figs3}.
Beyond accuracy, faithfully reproducing the interatomic interactions in light-element systems such as carbon requires addressing nuclear quantum effects, which arise from the quantum mechanical nature of light nuclei and can significantly influence vibrational properties and structural stability. To account for these effects, we employ the quantum thermal bath (QTB) method \cite{Filleter2009,Wu2022a}, which incorporates nuclear quantum effects into classical MD simulations through a colored-noise Langevin thermostat, without the prohibitive cost of path-integral-based approaches. This enables a more physically faithful treatment of the microscopic dynamics during the simulated friction process. All relevant computational parameters are detailed in the Supplementary Materials.

\section{results and discussions}
\subsection{The atomic structure of the frictional interface}

\begin{figure*}[htbp]
\centering
\includegraphics[width=0.8\textwidth]{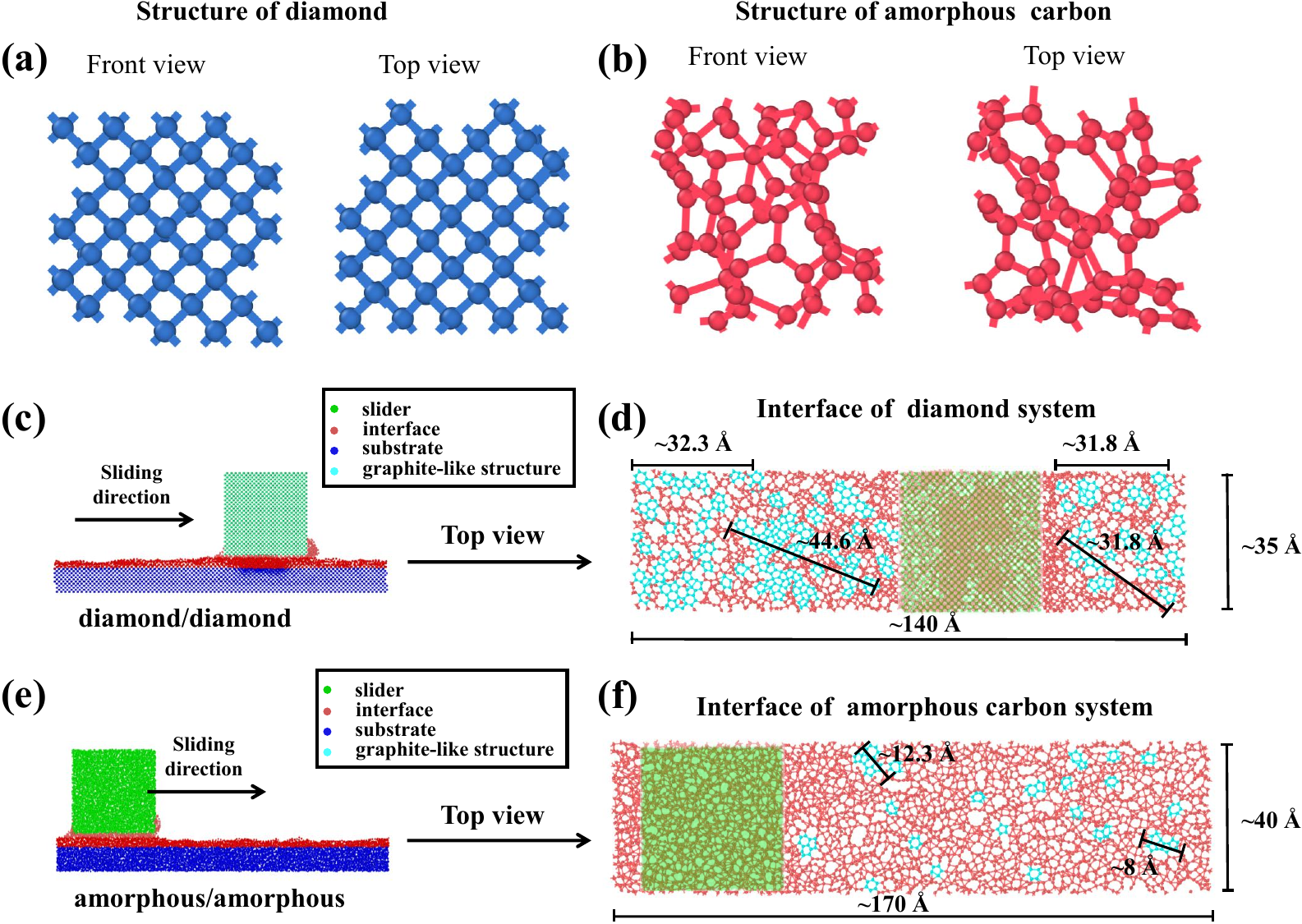}
\caption{Simulation models and friction-induced graphitization at carbon interfaces. Crystal structure of (a) diamond and (b) amorphous carbon used as the substrate materials. (c) Schematic of the diamond/diamond friction system and (d) the corresponding graphite-like structures (highlighted in light blue) formed at the frictional interface after sliding. (e) Schematic of the amorphous/amorphous carbon friction system and (f) the corresponding graphite-like structures (highlighted in light blue) formed at the frictional interface after sliding.}
\label{fig2}
\end{figure*}

As established in previous studies \cite{Li2020,Lin2022}, friction at carbon-based material interfaces can induce the formation of graphite-like structures. To illustrate the fundamental wear mechanism associated with this phenomenon, we present a schematic in FIG. \ref{fig1}. As shown in FIG. \ref{fig1}(a), two distinct local structural environments coexist at the graphitized frictional interface: regions dominated by graphite-like carbon rings, and regions where atoms retain stronger out-of-plane bonding to the substrate. The graphite-like regions, as depicted in FIG. \ref{fig1}(b), exhibit intrinsically weakened out-of-plane bonding, which renders the atoms within these regions susceptible to slipping in the friction process. In contrast, strongly bonded substrate regions anchor the surrounding atoms and act as structural reservoirs that continuously supply atoms to the interface during sliding. The interplay between these two types of regions governs the atomic-scale wear dynamics, as schematically illustrated in FIG. \ref{fig1}(c): atoms anchored in strongly bonded substrate regions undergo dynamic bond-breaking under frictional loading, migrate onto the graphite-like regions where out-of-plane bonding is weak, slip along the interface, and may subsequently re-bond to adjacent substrate sites. This cyclic process of bond breaking, lateral slip, and re-bonding constitutes the elementary atomic wear event at the carbon interface. Consequently, the characteristic length scale of the graphite-like reconstruction layer critically governs the efficiency of this process, and thereby the macroscopic wear rate.

To investigate the atomic-scale mechanisms underlying the wear behaviors of the two carbon systems, we construct two types of substrates representing the archetypal carbon bonding configurations: diamond (FIG. \ref{fig2}(a)) and amorphous carbon (FIG. \ref{fig2}(b)). A slider-on-substrate model is then employed, as illustrated in FIG. \ref{fig2}(c) and 2(e). In this model, a rigid slider moves over the substrate at a uniform velocity, while periodic boundary conditions are applied along the sliding direction to emulate friction on an effectively infinite substrate surface. The interfacial separation is set to approximately 3 Å, consistent with experimental observations that carbon atoms tend to establish microscopic contact at this length scale \cite{Wang2000,Rudenko2011}. The substrate dimensions are 140 Å × 35 Å for the diamond system and 170 Å × 40 Å for the amorphous carbon system. All friction simulations are performed at 600 K with a sliding velocity of 1 Å/ps. 

\begin{figure*}[htbp]
\centering
\includegraphics[width=0.8\textwidth]{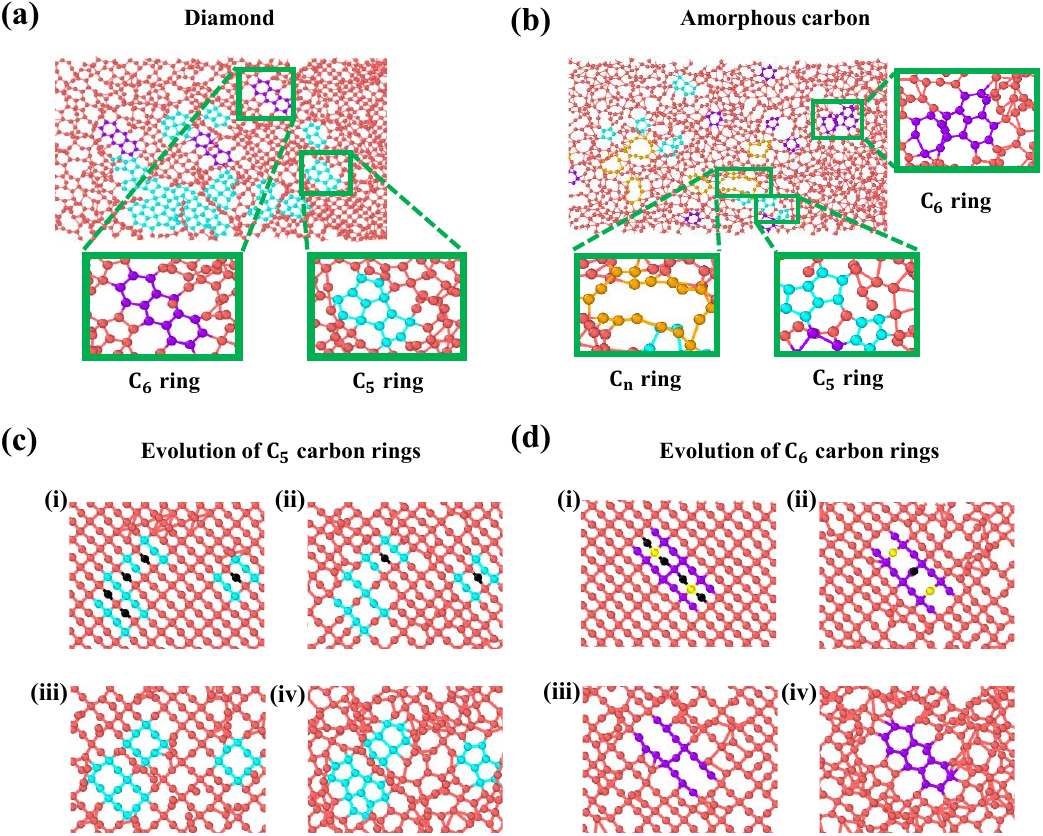}
\caption{Top view of the interfacial carbon layer formed at the (a) diamond/diamond and (b) amorphous/amorphous carbon frictional interfaces after sliding. Atoms are color-coded according to their ring-forming tendency: light blue atoms represent those that participate in the formation of five-membered carbon rings (\ce{C5}) and their transitional configurations; purple atoms represent those associated with six-membered carbon rings (\ce{C6}) and their transitional configurations; yellow atoms represent those involved in other n-membered carbon rings (Cn) and their transitional configurations. Dynamic evolution of (c) \ce{C5} and (d) \ce{C6} carbon rings at the diamond/diamond frictional interface during friction.}
\label{fig3}
\end{figure*} 

During friction, atoms at the diamond/diamond interface are progressively removed, leading to the formation of a disordered interfacial carbon layer approximately 4 Å thick (red region in FIG. \ref{fig2}(c)). The formation of this layer, consistent with experimental observations \cite{Liu2025}, confirms that our simulations faithfully capture the key interfacial evolution process. As shown in FIG. \ref{fig2}(d) and FIG. \ref{fig3}(a), within this interfacial carbon layer, a substantial number of regular 2D carbon rings (including hexagonal carbon rings and pentagonal carbon rings) form extensively connected networks. We define the characteristic length scale of the sp\textsuperscript{2} reconstruction layer as the maximum lateral extent of these continuously connected ring networks; at the diamond interface, this length scale exceeds 30~\AA, with the largest continuous network reaching up to 44.6~\AA. These graphite-like carbon rings are likely the atomic-scale origin of the experimentally observed graphitization at the diamond friction interface \cite{Liu2025,Wu2022}. In FIG. \ref{fig2}(f) and FIG. \ref{fig3}(b), the interfacial carbon layer at the amorphous/amorphous carbon interface contains fewer graphite-like carbon rings, which are sparsely distributed. The longest continuous graphite-like network in the amorphous carbon system is only about 12.3~\AA\ in length, giving a characteristic length scale nearly four times smaller than that of the diamond interface. Therefore, the spatial extent of graphitization at the amorphous interface is significantly smaller than that at the diamond interface after friction. 

To elucidate the formation pathway of these graphite-like structures, we trace the atomic trajectories throughout the sliding process, with the dynamic evolution of hexagonal (\ce{C6}) and pentagonal (\ce{C5}) carbon ring populations at the diamond/diamond interface presented in FIG. \ref{fig3}(c) and FIG. \ref{fig3}(d) respectively. The analysis reveals that the graphite-like structures at the diamond interface originate directly from the crystalline structure of the diamond substrate [FIG. \ref{fig2}(a)]. During friction, face-centered atoms within the subsurface diamond unit cells are selectively peeled away from the substrate, giving rise to an initial population of hollow square carbon rings at the interface. As sliding continues, further removal of carbon atoms at the shared boundaries between neighboring hollow squares causes these rings to form rectangular carbon ring configurations. These hollow square and rectangular carbon rings constitute the key intermediate structural states of the evolving interfacial carbon layer in the diamond/diamond friction system, as shown in the temporal snapshots of FIG. \ref{fig3}(c) and FIG. \ref{fig3}(d).

\begin{figure*}[htbp]
\centering
\includegraphics[width=0.8\textwidth]{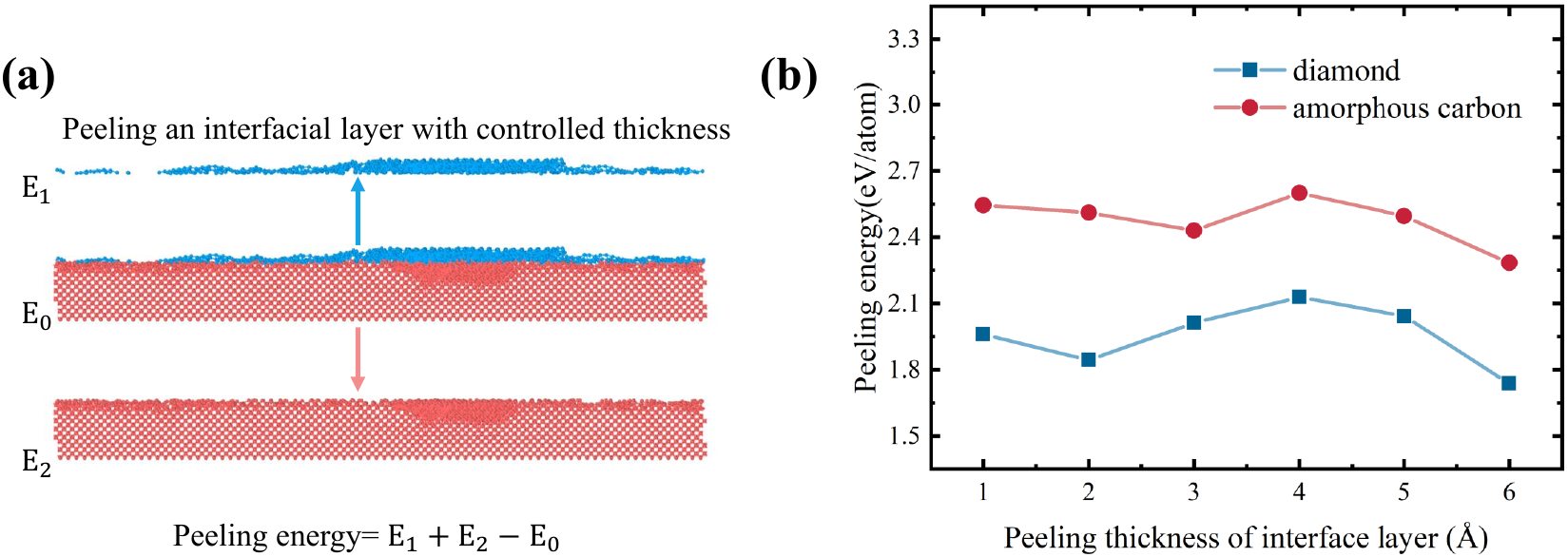}
\caption{Peeling energy analysis at carbon frictional interfaces. (a) Schematic definition of the peeling energy. (b) Peeling energy as a function of the thickness of interfacial carbon in diamond/diamond (red) and amorphous/amorphous carbon (blue) systems.}
\label{fig4}
\end{figure*} 

Upon continued sliding, the rectangular carbon ring intermediates can further relax and reorganize into continuous arrays of \ce{C6} carbon rings, as evident in FIG. \ref{fig3}(a) and the ring evolution statistics in FIG. \ref{fig3}(d). The formation of extended C6 networks is particularly favorable for lubrication, as the regular hexagonal lattice maximizes in-plane bonding while minimizing out-of-plane constraints on adjacent atoms \cite{Liu2021,Yan2024}. However, because the rectangular intermediate configuration requires a precise local atomic arrangement that is only accessible under specific geometric conditions at the diamond surface, the fully developed C6 network is not the predominant product. Instead, \ce{C5} carbon rings are the most abundant graphite-like species at the diamond interface (FIG. \ref{fig3}(a)). They evolve more readily from the hollow square intermediates through a lower-barrier structural rearrangement, which is shown in FIG. \ref{fig3}(c). Their spatial arrangement bears a close structural resemblance to penta-graphene \cite{Ewels2015}, and DFT calculations confirm that these \ce{C5} rings retain substantial sp2 hybridization character \cite{Liu2025}, endowing them with the weakened out-of-plane bonding that underlies their lubrication efficacy \cite{Yan2024,Zhang2024}. Collectively, both C5 and C6 rings contribute to building the extended, continuously connected graphite-like network that characterizes the diamond/diamond frictional interface and ultimately governs its superior wear resistance.

In contrast to the diamond system, the amorphous carbon interface exhibits a markedly different graphitization character, as revealed by the top view of its 4 Å-thick interfacial carbon layer after sliding (FIG. \ref{fig3}(b)). Additional atomic configurations during and after friction are provided in  FIG.~\ref{figs5}. While \ce{C5} and \ce{C6} carbon rings are present at this interface as well, the layer is additionally populated by large-cavity ring structures, \ce{Cn} (comprising ten or more carbon atoms), which are entirely absent at the diamond/diamond interface. These \ce{Cn} rings come from the disordered structure of the amorphous substrate. This disorder prevents the formation of the crystal-guided peeling pathway that produces \ce{C5} and \ce{C6} rings in diamond. During sliding, the pre-existing \ce{C5} and \ce{C6} rings at the amorphous interface remain structurally stable and resist bond fracture, consistent with experimental observations that graphitization at amorphous carbon interfaces is largely irreversible \cite{Loh2013,Wei2024}. Furthermore, disordered carbon atoms surrounding these graphite-like rings progressively reorganize into additional \ce{C5} and \ce{C6} configurations, generating small, locally ordered graphite-like structures \cite{SanchezLopez2003,Erdemir1996,Liu1996,Liu1997}. These locally ordered regions partially reduce interfacial friction \cite{Li2020,Ma2011}. However, their growth is severely curtailed by the surrounding \ce{Cn} rings, which are structurally rigid and resist transformation into \ce{C5} or \ce{C6} configurations even upon structural relaxation. As a result, the Cn rings act as topological barriers that interrupt the lateral connectivity of the graphite-like network and suppress the formation of large-scale continuous structures. Consequently, unlike the extended and interconnected graphite-like network at the diamond interface, the amorphous carbon interface retains a fragmented and spatially disordered distribution of graphite-like domains. This structural characteristic, as discussed below, directly underlies its inferior wear resistance.

The disparity in characteristic length scale between the two interfaces has direct consequences for interfacial bonding and atomic peeling: the extended reconstruction layer in diamond versus the isolated short-range patches in amorphous carbon produce fundamentally different bonding environments at the interface. Within graphite-like structures, carbon atoms form strong covalent bonds in the basal plane, yet their out-of-plane interactions with neighboring atoms are considerably weaker. This inherent bonding anisotropy controls how readily individual atoms or layers can be peeled under frictional loading. To quantify this anisotropy and compare it between the two systems, we calculate the energy required to peel successive interfacial carbon layers, as presented in FIG. \ref{fig4}. The peeling energy per atom in the diamond system is lower than that in the amorphous carbon system. Importantly, even at this lower value, the peeling energy remains substantial ($\sim$ 2 eV/atom), indicating that the continuous graphite-like network is stably anchored to the diamond substrate rather than freely detachable as isolated fragments. In the amorphous carbon system, the absence of a continuous graphite-like network means that interfacial atoms are not collectively stabilized. Instead, a larger population of isolated, weakly anchored atoms is exposed at the interface, making them far more susceptible to peel and wear during sliding. The peeling energy analysis thus provides quantitative support for the conclusion that the characteristic length scale of the graphite-like reconstruction layer is the decisive structural parameter governing wear resistance in carbon-based interfaces.

\subsection{Slipping-induced interfacial wear mechanism}

\begin{figure*}[htbp]
\centering
\includegraphics[width=0.8\textwidth]{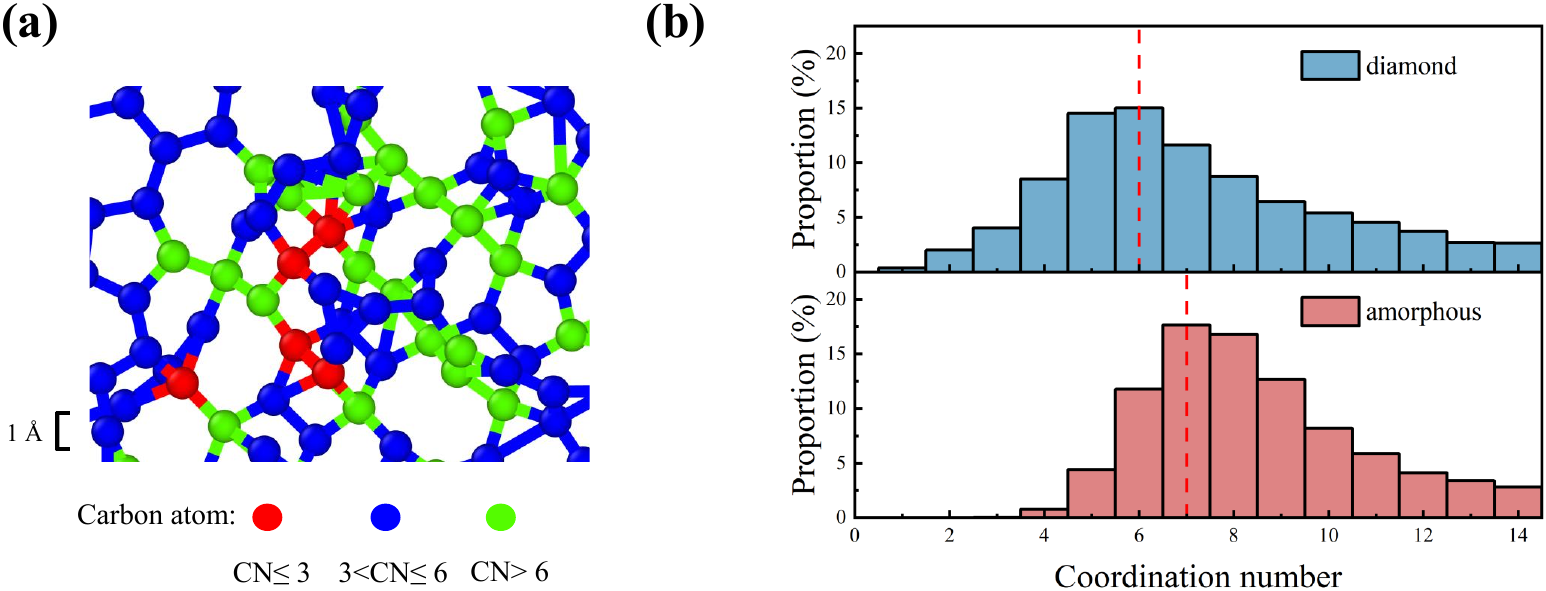}
\caption{Coordination number (CN) analysis of interfacial carbon atoms after friction. (a) Atomic visualization of interfacial carbon atoms in the diamond system, color-coded by their coordination number. (b) Statistical distribution of atomic coordination numbers at the frictional interface for the diamond (red) and amorphous carbon (blue) systems after sliding.}
\label{fig5}
\end{figure*} 

To elucidate how the contrasting graphitization characteristics of the two interfaces influence the differences in wear rate, we analyze the atomic coordination environments and slip dynamics at the interfaces. As established above, the graphite-like carbon layer reduces the out-of-plane bonding constraints on interfacial atoms, leaving a population of under-coordinated, weakly anchored atoms susceptible to lateral mobilization under the friction process. The atomic coordination number (\ce{Cn}) is defined here as the number of faces of the Voronoi polyhedron \cite{Takacs1986}. This provides a physically transparent metric for identifying such atoms: carbon atoms forming graphite-like rings typically exhibit \ce{Cn} values between 3 and 6 \cite{Rud2014}, whereas atoms with \ce{Cn} $>$ 6 reside in highly delocalized bonding environments with weak net out-of-plane interactions \cite{Trindle2020}, and those with \ce{Cn} $<$ 3 are severely under-coordinated with very few neighbors \cite{Trindle2020,Rasul2011,Sun2025}. These populations represent atoms prone to peeling and sliding. As visualized in FIG. \ref{fig5}(a), these bonded atoms are spatially distributed across the diamond interface. The comparative \ce{Cn} statistics in FIG. \ref{fig5}(b) reveal that the amorphous/amorphous interface hosts a markedly higher fraction of high-\ce{Cn} ($>6$) carbon atoms than the diamond/diamond interface. This distribution directly reflects the small characteristic length scale of the sp\textsuperscript{2} reconstruction layer at the amorphous interface, which is insufficient to collectively stabilize interfacial atoms through extended in-plane bonding, and accounts for the greater susceptibility of amorphous carbon to interfacial wear.

\begin{figure}[b]
\centering
\includegraphics[width=0.8\columnwidth]{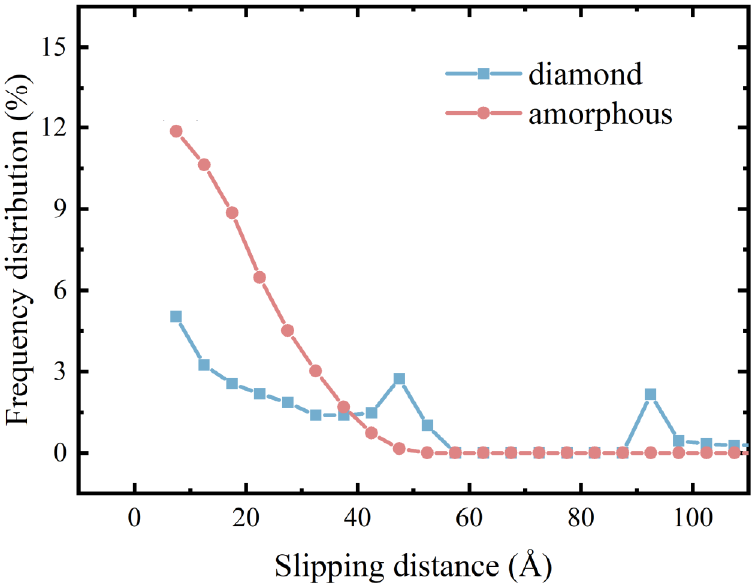}
\caption{Statistical distribution of atomic slip distances at the diamond/diamond (red) and amorphous/amorphous carbon (blue) frictional interfaces accumulated over a sliding duration of 0.05 ns.}
\label{fig6}
\end{figure}

The consequences of this difference in atomic bonding are directly manifested in the atomic slip statistics. As shown in FIG. \ref{fig6}, a large proportion of interfacial atoms in the amorphous carbon system accumulate slip distances of 15–25 Å within 0.05ns. This is because the sp\textsuperscript{2} reconstruction patches at the amorphous interface have a characteristic length scale of only $\sim$12~\AA. Atoms that become peeled at the boundaries of these domains can only slide over a short distance before being removed, leading to frequent short-range slip events and a high steady-state wear rate. In the diamond/diamond system, by contrast, far fewer atoms slip over this short distance range. A small proportion of atoms undergoes markedly longer slip events, extending to 45 Å and 95 Å. These distances closely match the lengths of the continuous graphite-like networks at the diamond interface (FIG. \ref{fig2}(d)). Detailed statistical results are provided in  FIG.~\ref{figs6}. These long-range slip events, though infrequent, are associated with substantial atomic rearrangement and localized interface roughening, as documented in  FIG.~\ref{figs4}. Nevertheless, because the majority of interfacial atoms in diamond are collectively stabilized within the extended graphite-like network, the overall rate of atom removal remains low, yielding macroscopically
superior wear resistance relative to amorphous carbon.

The mechanistic insights established above provide concrete guidance for engineering the graphitization behavior of carbon-based tribological coatings. For diamond-based systems, since the continuous graphite-like network at the frictional interface originates from the selective peeling of face-centered atoms within the subsurface diamond unit cells, one effective strategy to regulate the characteristic length scale of friction-induced graphitization is to preferentially protect these crystallographic sites. This can be achieved by engineering diamond surfaces to expose and stabilize the (111) close-packed planes to resist the peeling of face-centered atoms that initiates graphitization. This concept is analogous to the well-established preferential etching of crystalline silicon in alkaline solutions \cite{Barrio2012}, where the (111) planes are selectively preserved owing to their high atomic packing density and reduced chemical reactivity. Protecting the (111) planes in diamond coatings allows us to control the onset and spread of friction-induced graphitization, offering a way to tailor wear resistance for different tribological conditions. For amorphous carbon systems, which lack the long-range crystalline order necessary to develop a large characteristic length scale in the graphite-like reconstruction layer, an alternative route to improving tribological performance is through elemental doping. Introducing foreign atoms into the amorphous carbon system can promote the local formation of additional graphite-like structures and simultaneously modify the frictional properties of the interface \cite{Gao2005,Kim2006,Liu1996a}, thereby partially compensating for the structural limitations imposed by the absence of crystalline order. Crystallographic orientation control in diamond and compositional modification in amorphous carbon illustrate how the atomic-scale understanding of graphitization developed in this work can be translated into actionable design principles for next-generation wear-resistant carbon coatings.

\section{conclusions}
In this work, we develop a Deep Potential machine learning potential for diamond and amorphous carbon systems and employ it in large-scale MD simulations to investigate the atomic-scale mechanisms of friction-induced wear. This potential achieves DFT-level accuracy in describing atomic forces and energies, and correctly reproduces the experimentally known diamond (001) surface reconstruction. This provides a key benchmark for testing its ability to capture the evolution of atomic structures at interfaces.

Our simulations reveal that both diamond and amorphous carbon interfaces undergo graphitization under friction, but with fundamentally distinct structural characteristics. At the diamond/diamond interface, friction selectively peels face-centered atoms from subsurface unit cells. This generates hollow square and rectangular carbon ring intermediates. These intermediates then reorganize into extensive, continuously connected networks of \ce{C5} and \ce{C6} graphite-like rings, yielding a characteristic length scale of up to 44.6~\AA. At the amorphous/amorphous carbon interface, the lack of crystalline order prevents this structure-directed pathway. Instead, the interfacial carbon layer is interspersed with large-cavity \ce{Cn} carbon rings, which act as topological barriers that interrupt network connectivity and confine the characteristic length scale of graphite-like reconstruction to only $\sim$12.3~\AA.

The characteristic length scale of the graphite-like reconstruction layer governs the population of weakly bonded interfacial atoms, which is evidenced by the higher fraction of high-coordination-number (\ce{Cn} $>$ 6) atoms at the amorphous carbon interface. These atoms, in turn, determine atomic mobility under frictional loading. The amorphous carbon system exhibits abundant short-range atomic slip events (15–25 Å), reflecting the frequent peeling of atoms from the edges of isolated short-range sp\textsuperscript{2} patches and driving a higher wear rate. In diamond, the extended graphite-like reconstruction layer collectively stabilizes interfacial atoms, suppressing short-range detachment. The infrequent long-range slip events can reach up to 45 Å and 95 Å. These distances are comparable to the characteristic length scale of the graphite-like reconstruction layer, which confirms a direct structural origin for these slip events. Peeling energy calculations further corroborate these findings: the continuous graphite-like layer in diamond is stably anchored at $\sim$2~eV/atom, reducing the population of freely detachable isolated atoms and conferring superior wear resistance. These results establish a complete causal chain from the characteristic length scale of graphitization to isolated atom population to macroscopic wear rate.

Building on this mechanistic understanding, we believe that wear resistance in diamond coatings can be actively engineered by protecting the (111) close-packed crystallographic planes, which suppresses the initiation of the graphitization pathway and allows the characteristic length scale of the reconstruction layer to be tuned for specific tribological applications. For amorphous carbon, elemental doping offers a complementary route to promote local graphite-like ordering and improve tribological performance. More broadly, this work demonstrates that machine learning potential-driven MD simulations are a powerful and generalizable tool for bridging atomic-scale structural evolution and macroscopic tribological performance. This approach is directly applicable to other technologically important interface systems, including 2D layered materials, nitride coatings, and multi-component amorphous films, where the interplay between local bonding, structural order, and wear remains incompletely understood.

\begin{acknowledgments}
This work is supported by the Strategic Priority Research Program of the Chinese Academy of Sciences (XDB0470103), the National Natural Science Foundation of China (U24A20113, 52471104), Zhejiang Provincial Natural Science Foundation of China (LD24E010002), China Postdoctoral Science Foundation (2023M743630), and the "Innovation Yongjiang 2035" Key R\&D Program (2024Z095, 2024Z138).
\end{acknowledgments}

\nocite{*}
\bibliographystyle{apsrev4-2}
\bibliography{apssamp}

\clearpage
\onecolumngrid

\begin{center}
{\large\bfseries Supplementary Materials: Role of Characteristic Length Scale in Interface Graphitization-Induced Wear Resistance of Diamond and Amorphous Carbon}
\end{center}
\vspace{1em}

\setcounter{figure}{0}
\setcounter{table}{0}
\setcounter{section}{0}
\setcounter{equation}{0}
\renewcommand{\thefigure}{S\arabic{figure}}
\renewcommand{\thetable}{S\arabic{table}}
\renewcommand{\thesection}{S\arabic{section}}
\renewcommand{\theequation}{S\arabic{equation}}
\renewcommand{\theHfigure}{S\arabic{figure}}
\renewcommand{\theHtable}{S\arabic{table}}
\renewcommand{\theHsection}{S\arabic{section}}
\renewcommand{\theHequation}{S\arabic{equation}}

\section{Details of the DP model construction and testing}
Empirical models of atomic interaction are conventionally used by conventional MD, which can lead to the unsatisfactory accuracy of MD simulations due to their simplistic analytical forms. DP method can generate a potential model that achieves first-principles computational accuracy through sampling a series of accurate DFT data. Our DP training database contains the configurations of diamond, graphite, amorphous carbon, and the (001) reconstructed surface of diamond. To obtain these accurate training data, we use the Vienna ab initio simulation package (VASP) with Perdew-Burke-Ernzerh of exchange-correlation functional. We use a plane-wave cutoff energy of 500 eV in the structural relaxation calculations, and KSPACING= 0.1 is set to converge energy and atomic force. In order to explore more atomic configurations under different temperature and pressure, the MD simulations in the DP training step are carried out by the LAMMPS code with periodic boundary conditions. We use the isobaric-isothermal (NPT) ensemble with temperature set from 10 K to 1200 K and pressure set from 1 bar to 5 GPa to sample configurations.

\begin{figure}[htbp]
	\centering
	\includegraphics[width=0.75\textwidth]{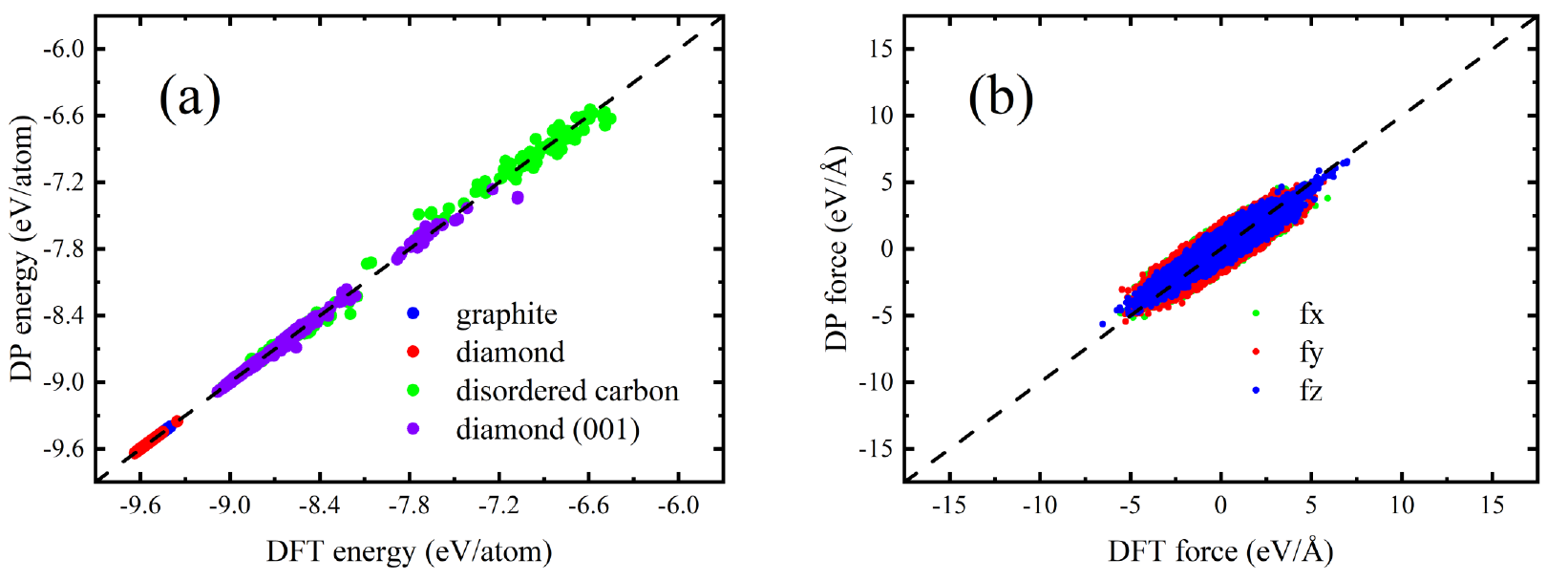}
	\caption{Accuracy testing of the DP model. Comparison of energies (a) and atomic forces (b) using DP model and training datasets consist of DFT data (including crystalline graphite, diamond, diamond (001) surface and disordered carbon). Our training dataset contains more 163803 DFT results (41.73 \% diamond, 6.35 \% graphite, 21.20 \% diamond (001) surface, 30.73 \% disordered carbon). Our model's RMSE for the energies of all involved structures is 43 meV/atom.}
	\label{figs1}
	\end{figure}

To construct a test dataset that is independent of training set, we use 0.15 ns MD simulations (300 K, isobaric-isothermal ensemble) to produce structural evolution trajectories of diamond structures, diamond (001) surfaces (initial structure is obtained from ab initio molecular dynamics simulations at 300 K with a time scale of 1000 fs.), and disordered carbon structures (initial structures can be obtained from MD simulations of bulk diamond from 3000 K to 300 K) with 64 atoms. We randomly select 1000 structures from each trajectory sample to form a testing set to ensure structures included in this testing set are independent of the training set.

\section{Details of the MD simulation setup}
We construct a diamond/diamond friction system and an amorphous/amorphous carbon friction system. The diamond/diamond system comprising a cubic diamond slider (green part in Fig. 1(a), rigid, 8000 atoms, approximately 35 \AA$\times$35 \AA$\times$35 \AA) and a diamond substrate (blue part in Fig. 1(a), 12,800 atoms, approximately 140 \AA$\times$35 \AA$\times$12.5 \AA). The amorphous/amorphous carbon friction system composed of a rigid amorphous carbon slider (green part in Fig. 1(b). 8000 atoms, ~40 \AA $\times$ 40 \AA $\times$ 40 \AA) and an amorphous carbon substrate (blue part in Fig. 1(b), 12,800 atoms, ~170 \AA $\times$ 40 \AA $\times$ 17 \AA).

When the slider's dimensions are significantly smaller than those of the substrate, atomic-scale interactions at the interface dominate the frictional behavior. Under such conditions, the contributions from the slider's flexibility and elastic deformation are typically negligible. Consequently, it is frequently adopted in theoretical modeling to initially treat the slider as a rigid entity. Therefore, our work corresponds to this physical scenario, aligning with the foundational assumptions prevalent in many classical studies of tip-based sliding. Given that our amorphous structure is obtained through rapid supercooling (using the NPT ensemble by cooling from 4000K to 300K at a rate of 370K/ps), maintaining a specific tip geometry (e.g., conical or spherical) is challenging; thus, a cubic shape is selected for its simplicity and reproducibility. The primary objective of the current work is to clearly reveal the atomic processes at the sliding interface. The rigid slider assumption provides a clean, well-controlled theoretical framework essential for isolating and analyzing these fundamental mechanisms. For the QTB method, it is necessary to couple it with either an NVE or an NPH ensemble. In our simulation, an NVE ensemble is applied to the slider. This choice also renders the slider effectively rigid (an NPH ensemble can lead to instability due to its excessively exposed surfaces). For the substrate, an NPH ensemble is employed, with the thermostat/barostat controller set to iso 1 1 0.5. Regarding the QTB parameters, the most critical are the maximum vibrational frequency, the number of frequency bins, and the damping coefficient, which are set to 50 (THz), 200, and 1 (1/ps), respectively. Detailed definitions of these parameters can be found in the LAMMPS documentation. And molecular dynamics software package used in our work are DP-GEN and DeePMD-kit. In the QTB method, temperature is defined through the kinetic energy of the particles. The friction force is defined as the sum of the forces on all atoms in the slider in the direction opposite to its motion.

\begin{figure}[htbp]
\centering
\includegraphics[width=0.75\textwidth]{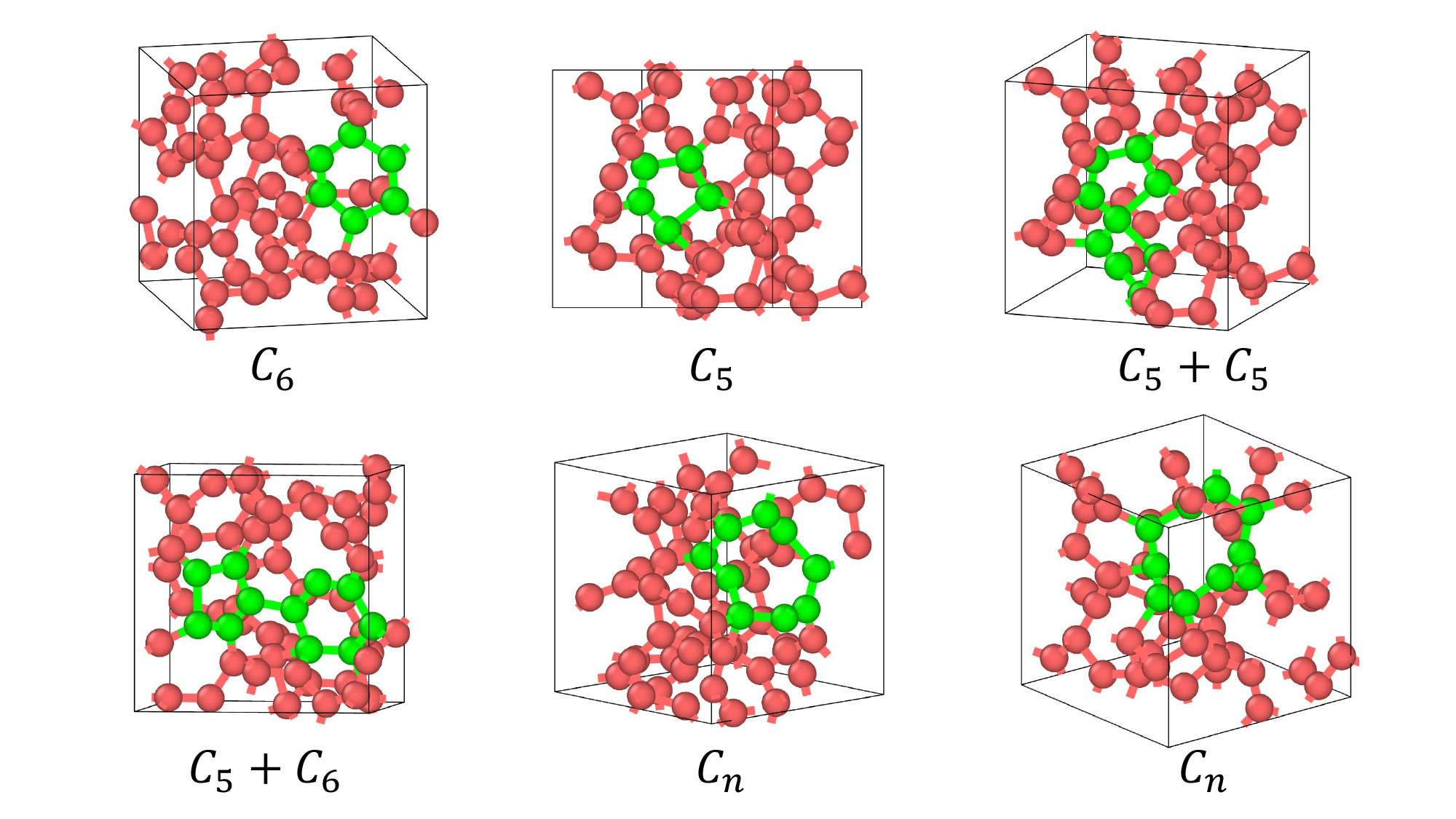}
\caption{Some representative disordered carbon structures in the testing set. The testing set includes several disordered carbon structures that involve the evolution of local C5, C6, and Cn carbon rings.}
\label{figs2}
\end{figure}

\begin{figure}[htbp]
\centering
\includegraphics[width=0.75\textwidth]{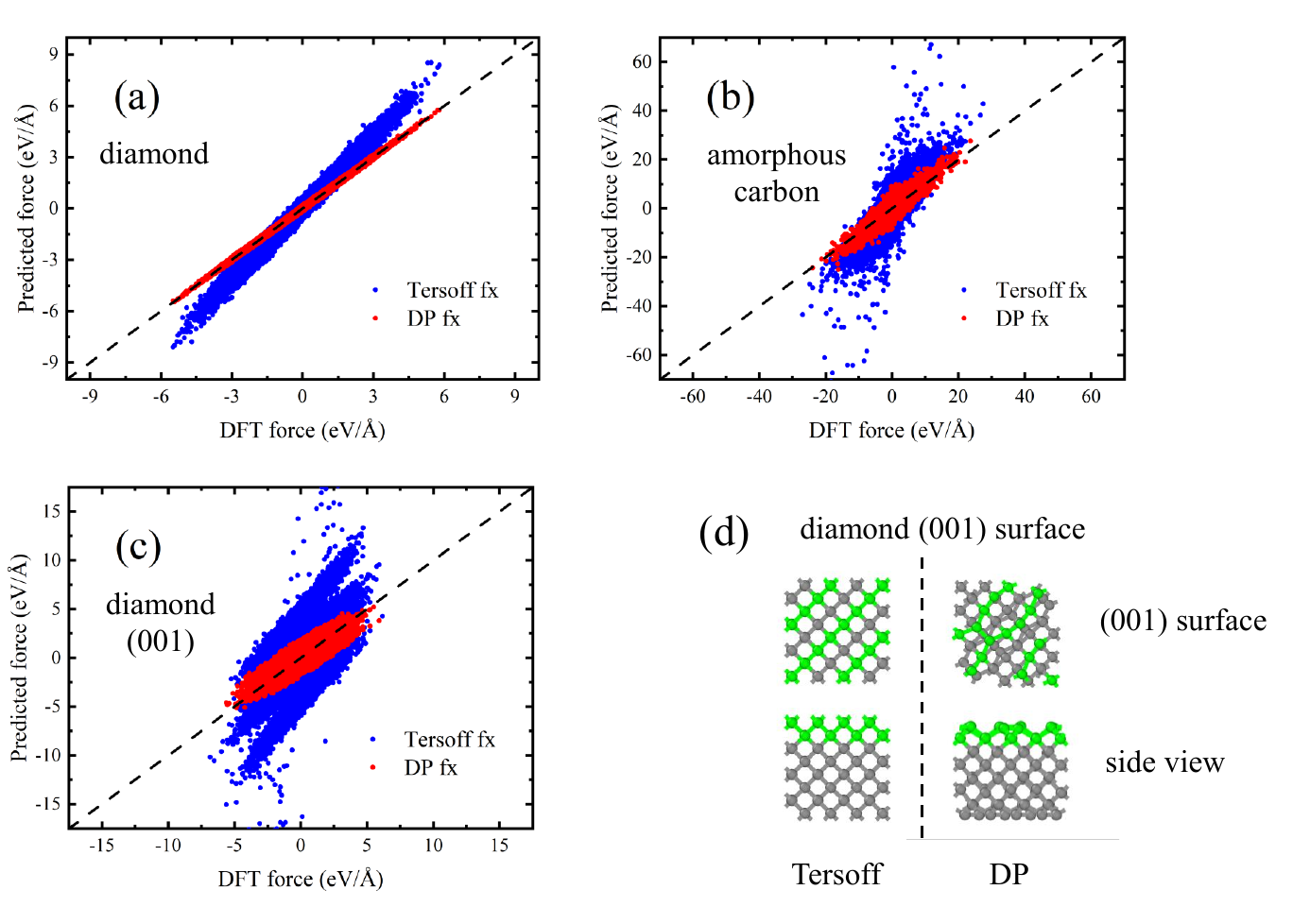}
\caption{Results of model testing. Comparison of atomic forces calculated using DP model and Tersoff potential based on testing datasets (including bulk diamond (a), disordered carbon (b) and diamond (001) surface (c). The left and right sides of (d) depict the diamond (001) surface reconstruction obtained through the Tersoff potential and DP models, respectively.}
\label{figs3}
\end{figure}

\begin{figure}[htbp]
\centering
\includegraphics[width=0.75\textwidth]{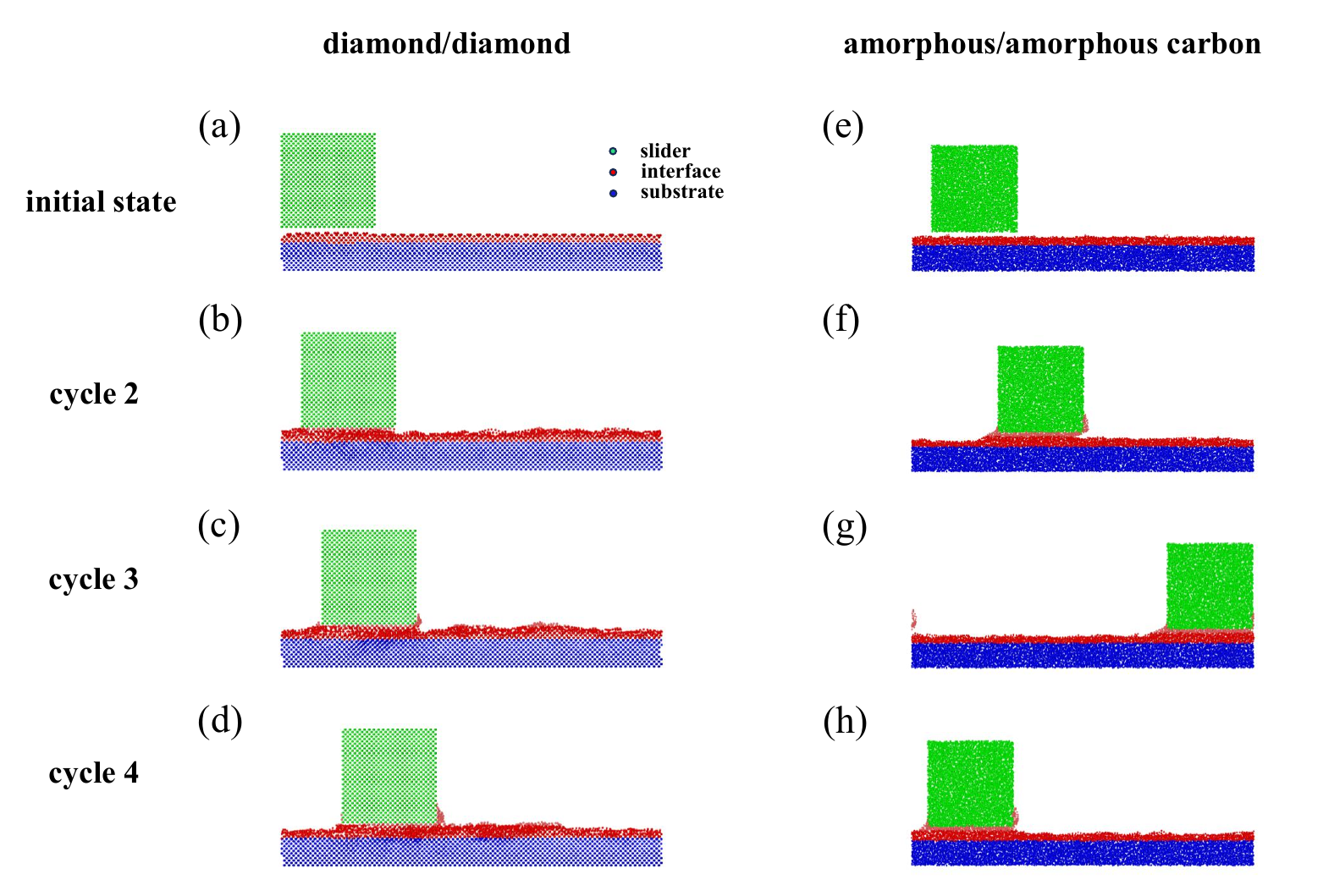}
\caption{(a)-(d) The interface carbon layer at the diamond/diamond interface undergoes significant morphological changes during 4-cycle friction, indicating the occurrence of localized detachment and sliding. (e)-(h) The interface carbon layer at the amorphous/amorphous carbon system exhibits minimal morphological changes during 4-cycle friction, suggesting that localized detachment is not pronounced.}
\label{figs4}
\end{figure}

\begin{figure}[htbp]
\centering
\includegraphics[width=0.95\textwidth]{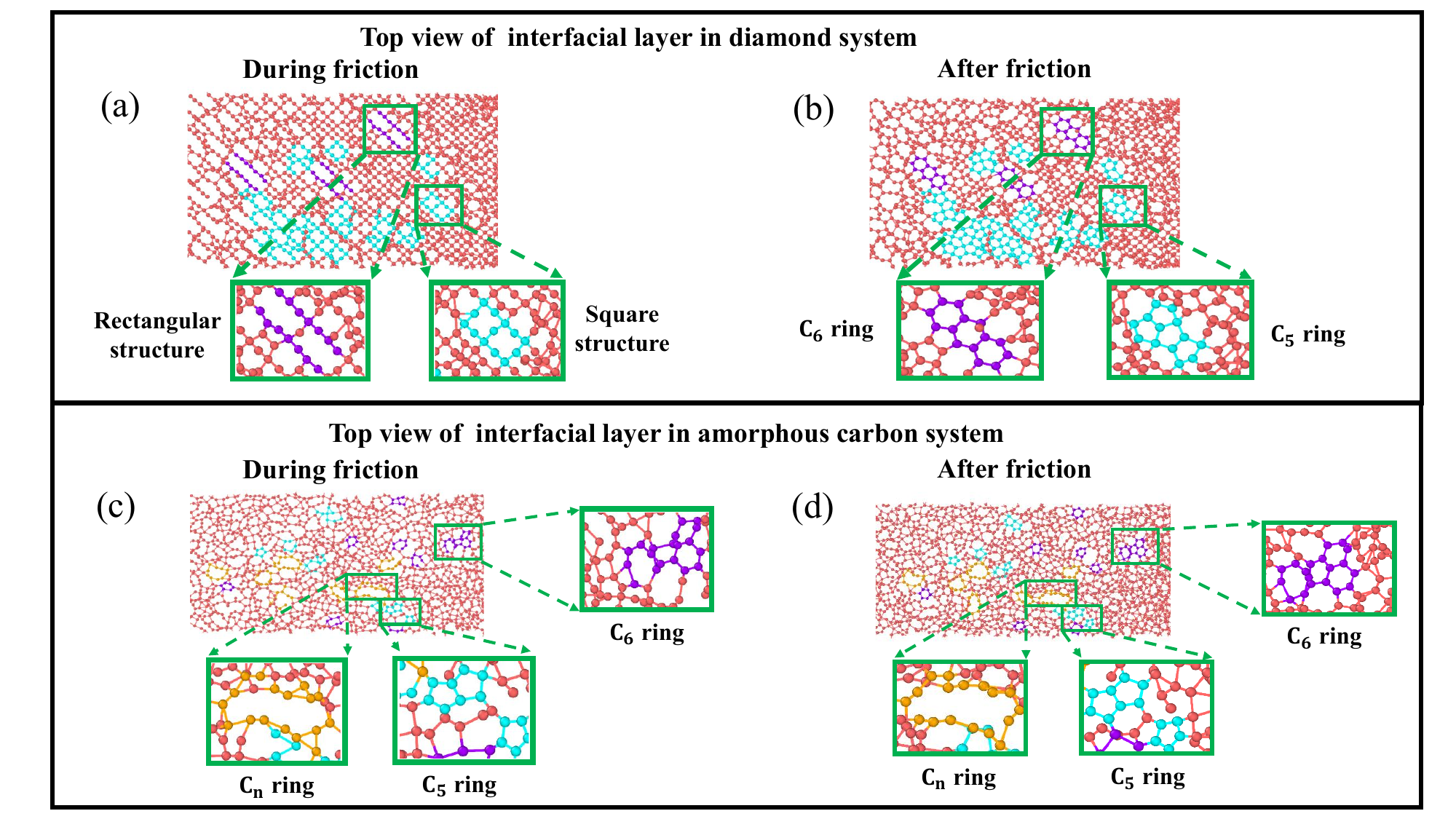}
\caption{Top view of interfacial carbon layer (4 \AA thickness) at the diamond/diamond (a), (b) and amorphous/amorphous (c), (d) frictional interface after friction. Atoms marked in light blue represent those that can transform into C5 carbon rings and their transitional configurations. Atoms marked in purple represent those that can transform into C6 carbon rings and their transitional configurations. Atoms marked in yellow represent those that can transform into Cn carbon rings and their transitional configurations.}
\label{figs5}
\end{figure}

\begin{figure}[htbp]
\centering
\includegraphics[width=0.95\textwidth]{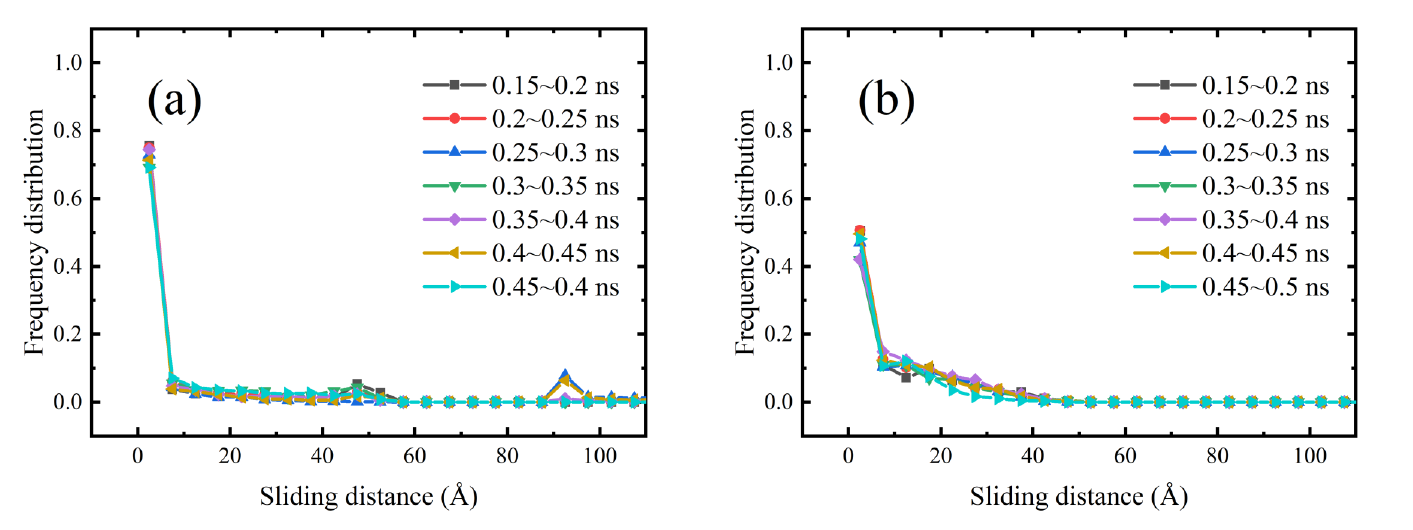}
\caption{Frequency distribution of atomic slip distances per 0.05 ns interval in diamond (a) and amorphous carbon (b) systems.}
\label{figs6}
\end{figure}

\begin{table}[htbp]
	\centering
	\caption{The number of atoms within the amorphous interfacial layer region (90 \AA $<$ 105 \AA) before and after a single friction event during the time interval of 015 ns to 0.33 ns (cycle 2- cycle 3).}
	\label{tab:system_properties}
	\begin{ruledtabular}
	\begin{tabular}{ccccccccccc}
	\textrm{System} & \textrm{0.15} & \textrm{0.17} & \textrm{0.19} & \textrm{0.21} &\textrm{0.23} & \textrm{0.25} & \textrm{0.27} & \textrm{0.29} & \textrm{0.31} &\textrm{0.33} \\
	\colrule
	diamond   & 226 & 226 & 227 & \textbackslash & \textbackslash & \textbackslash &196 & 197 & 196 & 194 \\
	amorphous & 223 & 221 & 227 & 229 & \textbackslash & \textbackslash &\textbackslash & 220 & 229 & 226 \\
	\end{tabular}%
	\end{ruledtabular}
	\end{table}


\end{document}